\documentstyle[12pt,aasms4,psfig]{article}
\newcommand{\Ha}{H$\alpha$}
\newcommand{\Lm}{$L_{\rm{mech}}$}
\newcommand{\nhi}{$N(\rm{H}^\circ)$}
\newcommand{\nhei}{$N(\rm{He}^\circ)$}
\newcommand{\nheii}{$N(\rm{He}^+)$}
\newcommand{\ovi}{O~VI $\lambda$1035}
\newcommand{\cii}{C~II $\lambda$1335}
\newcommand{\ciia}{C~II $\lambda$1036}
\newcommand{\nv}{N~V $\lambda$1240}
\newcommand{\siiv}{Si~IV $\lambda$1400}
\newcommand{\civ}{C~IV $\lambda$1550}
\newcommand{\heii}{He~II $\lambda$1640}
\newcommand{\niv}{N~IV $\lambda$1720}
\newcommand{\Hb}{H$\beta$}
\newcommand{\Brg}{Br$\gamma$}
\newcommand{\Pab}{Pa$\beta$}
\newcommand{\Myr}{M$_{\odot}$~yr$^{-1}$}
\newcommand{\Ms}{M$_{\odot}$}
\newcommand{\Mup}{$M_{\rm{up}}$}
\newcommand{\Mlow}{$M_{\rm{low}}$}

\newcommand{\Zs}{Z$_{\odot}$}

\newcommand{\Md}{$\dot{M}$}

\newcommand{\Teff}{$T_{\rm{eff}}$}
\tighten

\received{16 September 1998}
\accepted{4 January 1999}
\journalid{337}{15 January 1989}
\articleid{11}{14}

\slugcomment{Submitted to ApJS}

\begin{document}

\title{Starburst99:\\
 Synthesis Models for Galaxies with Active Star Formation}

\author{Claus Leitherer}
\affil{Space Telescope Science Institute\footnote{Operated
 by AURA for NASA under contract NAS5-26555}, 3700 San Martin Drive, Baltimore,
       MD 21218\\
       e-mail: leitherer@stsci.edu}

\author{Daniel Schaerer}
\affil{Observatoire Midi-Pyrenees,  14, Av. E. Belin, F-31400 Toulouse, France\\
       e-mail: schaerer@obs-mip.fr}

\author{Jeffrey D. Goldader}
\affil{Univ. of Pennsylvania, Dept. of Physics \& Astronomy, Philadelphia, 
       PA 19104-6396\\
       e-mail: jdgoldad@dept.physics.upenn.edu}

\author{Rosa M. Gonz\'alez Delgado}
\affil{Inst. Astrof\'{\i}sica de Andaluc\'{\i}a,
   Apartado 3004, E-18080 Granada, Spain\\
       e-mail: rosa@iaa.es}

\author{Carmelle Robert}
\affil{D\'epartement de Physique and Observatoire du mont M\'egantic,
            Universit\'e Laval, Qu\'ebec, QC, G1K 7P4, Canada\\
     e-mail: carobert@phy.ulaval.ca}

\author{Denis Foo Kune}
\affil{Macalester College, Dept. of Physics \& Astronomy, 1600 Grand Ave.,
       St. Paul, MN 55105\\
          e-mail: dfookune@macalester.edu\\
and\\
Space Telescope Science Institute, 3700 San Martin Drive, Baltimore,
       MD 21218}

\author{Du\'{\i}lia F. de Mello}       
\affil{Space Telescope Science Institute, 3700 San Martin Drive, Baltimore,
       MD 21218\\
       e-mail: demello@stsci.edu}

\author{Daniel Devost}       
\affil{Space Telescope Science Institute, 3700 San Martin Drive, Baltimore,
       MD 21218\\
and\\
D\'epartement de Physique and Observatoire du mont M\'egantic,
            Universit\'e Laval, Qu\'ebec, QC, G1K 7P4, Canada\\
       e-mail: devost@stsci.edu}       

\and

\author{Timothy M. Heckman}
\affil{Physics and Astronomy Department, Johns Hopkins University, Homewood
      Campus, Baltimore, MD 21218\\
      e-mail: heckman@pha.jhu.edu}         

\begin{abstract}

Starburst99 is a comprehensive set of model predictions for spectrophotometric 
and related properties of galaxies with active star formation. The models
are an improved and extended version of the data set previously published
by Leitherer \& Heckman (1995). We have upgraded our code by implementing 
the latest set of stellar evolution models of the Geneva group and the
model atmosphere grid compiled by Lejeune et al. (1997). Several 
predictions which were not included in the previous publication are shown 
here for the first time. The models are presented in a 
homogeneous way for five metallicities
between $Z$~=~0.040 and 0.001 and three choices of the initial mass function.
The age coverage is $10^6$ to $10^9$~yr. We
also show the spectral energy distributions which are used to compute colors
and other quantities. The full data set is available for 
retrieval at {\tt http://www.stsci.edu/science/starburst99/}.
This website allows users
to run specific models with non-standard parameters as well. We also make
the source code available to the community.

\end{abstract}


\section{Introduction}

Synthesis models for comparison with observed galaxy properties, and in 
particular with spectral energy distributions, have become increasingly popular
in recent years. This is partly due to improvements in stellar libraries and
models. An overview over the latest achievements can be found in the conference
volume edited by Leitherer, Fritze-v.~Alvensleben, \& Huchra (1996b). At least
equally important is the possibility of electronic dissemination of the data,
either by CD-ROM or via the internet. Most groups who provide such models to
the community make them available electronically. See Leitherer et al. (1996a)
for a compilation of a suite of models.

This paper concerns a particular set of synthesis models: models which are
optimized to reproduce properties of galaxies with active star formation. 
In the absence of an active galactic nucleus, most radiative properties of 
such galaxies are determined by their massive-star content. The most extreme
examples are referred to as starbursts (Weedman et al. 1981; 
Weedman 1987; Moorwood 1996) but the models in this
paper are often applicable to less extreme star-forming regions and
galaxies, like 30~Doradus and M33, as well. 
Leitherer \& Heckman (1995; hereafter LH95) published a homogeneous grid of
synthetic starburst models. These models, together with those of 
Bressan et al. and Bruzual \& Charlot (both published on the CD-ROM of 
Leitherer et al. 1996a), Fioc \& Rocca-Volmerange (1997), and 
Cervi\~no \& Mas-Hesse
(1994), are widely used to aide the interpretation of galaxy observations.

Although the LH95 models are still recently up-to-date and no major error
was found, we decided to generate and release a new generation
of models. There are two reasons for this. First, stellar modeling is 
continuously advancing, and some known shortcomings of the stellar atmosphere
and evolution models used in LH95 have been improved. Specifically, we have
included the new model atmosphere grid of Lejeune, Buser, \& Cuisinier (1997)
and the latest Geneva evolutionary tracks in our code. Other 
modifications include the possibility to perform isochrone synthesis, a
technique which was pioneered by Charlot \& Bruzual (1991).

The second motivation for an update to LH95 concerns the distribution method.
While the LH95 data set is unofficially available via our website, the 
principal publication is on paper. Many users found this inconvenient
and encouraged us to publish fully electronically. Therefore we decided not 
to include any tables with model predictions and only a subset of the figures
in the hardcopy version of this paper. The full data set with all the figures
is available in the electronic version of the journal  and via our webpage, 
which is an integral
part of this publication. The reader should visit the website at
{\tt http://www.stsci.edu/science/starburst99/}
to view the figures and to download the
tables. Moreover, this website offers the opportunity to run tailored models
remotely and to access our source code. The reader should use this paper
as a guide when visiting our webpage.

The organization of the paper is as follows: In \S~2 we outline
our computation technique, the model assumptions, and the improvements over
LH95. This section also explains how to reach our website and how to navigate.
The spectral energy distributions are presented in \S~3. Spectral line 
profiles in the ultraviolet (UV) are discussed in \S~4. In \S~5 we provide 
numbers for the stellar inventory. Luminosities and colors are in \S~6 and 7,
respectively. Properties related to the far-UV are presented in \S~8. Some
other useful diagnostic lines are in \S~9. \S~10 covers the mass and energy
return of a stellar population. We conclude with \S~11.

\section{Model assumptions and computational technique}

The present model set is an extension of our previous work which was published
in several papers. Most of the earlier tables are in LH95. The energy 
distributions used in LH95 were made available electronically in Leitherer et
al. (1996a) and on our website. 
Ultraviolet spectra between 1200~\AA\ and 1850~\AA\ at 0.75~\AA\ resolution
were originally published in Leitherer, Robert, \& Heckman (1995) and
distributed electronically in Leitherer et al. (1996a). A set of spectra
around \ovi\ and Ly$\beta$ was discussed by Gonz\'alez Delgado, Leitherer,
\& Heckman (1997). All mentioned models have been re-computed and included in
the present database.
The covered parameter space is similar to the one in LH95. Since arguments are
given in LH95 for the choice of these particular parameters, we will be brief
and only address those issues in more detail which are different from LH95.

We consider two limiting cases for the {\em star-formation law}: 
an instantaneous
burst of star formation and star formation proceeding continuously at a 
constant rate. The instantaneous models (also referred to as a single stellar
population) are normalized to a total mass of $10^6$~\Ms. The star-formation
rate of the continuous model is 1~\Myr. These normalizations were chosen
to produce properties which are typical for actual star-forming regions in
galaxies.

As we did in LH95, we offer three choices for the {\em stellar initial mass 
function} (IMF). The reference model is a 
power law with exponent $\alpha = 2.35$ between low-mass and high-mass
cut-off masses of \Mlow~=~1~\Ms\ and \Mup~=~100~\Ms, respectively. This 
approximates the classical Salpeter (1955) IMF. Most observations of 
star-forming and starburst regions are consistent with a Salpeter IMF ---
although the uncertainties can be large (Leitherer 1998; Scalo 1998). For
comparison, we also show the results for an IMF with $\alpha = 3.3$ between
1 and 100~\Ms, which has a higher proportion of low-mass stars. Finally,
we computed models with a truncated Salpeter IMF: $\alpha = 2.35$,
\Mlow~=~1~\Ms, \Mup~=~30~\Ms. The models can easily be re-normalized to other
\Mlow\ values. The lower mass cut-off 
is often only a scaling factor since low-mass stars have a
negligible contribution to the properties of the stellar population, except for
the mass locked into stars. Absolute quantities (such as, e.g., the number of
ionizing photons) can be rescaled to other \Mlow\ values by multiplying our
results by factors of 0.39, 1.00, 1.80, 3.24 for \Mlow~=~0.1, 1.0, 3.5, 
10~\Ms. Relative quantities (like colors) should, of course, not be scaled. 
Note that 10~\Ms\ stars are no longer negligible for many model
predictions, and the models should be recalculated for this cut-off mass, rather
than scaling the existing data. Common sense is necessary when applying this
scaling factor to instantaneous models in particular. The lifetime of a 2~\Ms\
star is only somewhat above 1~Gyr. Therefore burst models at 1~Gyr are actually
only represented by a population of 1~--~2~\Ms\ stars. Clearly even a small
change of the lower mass cut-off will drastically alter the model predictions.  
\footnote{A typographical error was introduced
in eq.~(3) of LH95: the numerator should be 0.80 instead of --0.35.}

We have implemented the new set of {\em stellar evolution models} of the Geneva
group --- as opposed to the Maeder (1990) models in LH95. For masses of
12~--25~\Ms\ (depending on metallicity)
and above we are using the tracks with enhanced mass loss of Meynet
et al. (1994). The tracks of Schaller et al. (1992), Schaerer et al. (1993a,  
1993b), and Charbonnel et al. (1993) with standard mass loss are used between
12 and 0.8~\Ms. We have favored the enhanced mass-loss models over the
standard ones because they are a better representation of most Wolf-Rayet (WR) 
star properties, except for the mass-loss rate (\Md) itself. The high mass-loss
rates of WR stars are in conflict with observations (Leitherer, Chapman, \&
Koribalski 1997). This may suggest the lack of one or another ingredient
in the evolution models, like rotation induced mixing processes (Maeder 1995).
The evolutionary tracks do not take into account binary evolution. Inclusion
of Roche-lobe overflow in binary systems can modify some of the predictions,
in particular during the WR phase. Models taking these effects into account
were computed by Cervi\~no, Mas-Hesse, \& Kunth (1997), Vanbeveren, 
Van~Bever, \& de Donder (1997), Schaerer \& Vacca (1998), and Dionne (1999).

The Geneva models include the early asymptotic-giant-branch (AGB) evolution 
until the first thermal pulse for masses $>$~1.7~\Ms. Stellar evolution
models in the AGB phase are rather dependent on input assumptions, and therefore
quite uncertain. A recent comparison between different sets of AGB models 
was made by Girardi \& Bertelli (1998).  

Five {\em metallicities} 
are available: $Z = 0.040$, 0.020 (=~\Zs), 0.008, 0.004, and
0.001. These are the metallicities of the evolutionary models. All element
and isotope ratios are independent of metallicity. Note that we do not treat
chemical evolution self-consistently: each stellar generation has the same
metallicity during the evolution of the population. The error introduced
by this simplification is negligible as long as star formation proceeds
for less than about 1~Gyr, otherwise self-consistent models are needed, such
as those by M\"oller, Fritze-v.~Alvensleben, \& Fricke (1997).

The models cover the {\em age range} $10^6$ to $10^9$~yr. In principle we
could have evolved the populations up to a Hubble time but we decided not to
do so. The evolutionary models are optimized for massive stars and become
less reliable for older ages. Many properties can still be safely  predicted
beyond 1~Gyr whereas others (such as, e.g., near-infrared colors) become 
increasingly uncertain. We will comment on uncertainties when the individual
figures are discussed. 

{\em Atmosphere models} in LH95 were from Kurucz (1992) for most stars, 
supplemented by the Schmutz, Leitherer, \& Gruenwald
(1992) models for stars with strong winds, and black bodies for the coolest
stars. Lejeune et al. (1997) have produced a new, homogeneous grid of
atmospheres covering the entire Hertzsprung-Russell diagram (HRD) populated
by the evolutionary models, including the coolest stars. This new grid has been
implemented in our synthesis code. We used the corrected grid whose fluxes
produce colors in agreement with empirical color-temperature relations.
The original atmosphere models do not
include metallicities of 0.008 and 0.004. We interpolated between adjacent
metallicities to obtain grids whose metallicities match those of the 
evolutionary models. No interpolation was required for 0.040, 0.020, and 0.001.
As in LH95, the Schmutz et al. (1992) extended model atmospheres are used for
stars with strong mass loss. The prescription for the switch between extended
and plane-parallel atmospheres is the same as in LH95. The extended 
atmospheres are assigned to all stars with surface temperature $>$25,000~K 
and hydrogen content on the surface $<$0.4. We decided not to
use extended model atmospheres for stars closer to the main-sequence since
the benefit of improved wind treatment would be more than compensated
by the limited metallicity range available for such models.

The {\em nebular continuum} has been added to all spectrophotometric 
quantities, except to the energy distributions in 
Figures~7a-e, 8a-e, 9a-e, 10a-e, 11a-e, and 12a-e. We
have assumed that all the stellar far-UV photons below 912~\AA\ are 
converted into free-free and free-bound photons at longer wavelengths (case B).
However, the stellar far-UV flux has {\em not} been removed in the spectral 
energy distributions, as this would probably not be desirable for the user.
To alert the user, the hydrogen and helium emission coefficients of Ferland 
(1980) do not extend longward of 4.5~$\mu$m. Therefore the nebular continuum
at these wavelengths is undefined in our code. 
In practice, this is hardly relevant since
dust emission often dominates at 
$\lambda > 3$~$\mu$m (Carico et al. 1988) and our models become inapplicable 
anyway.

All models shown here have been computed with the {\em isochrone synthesis}
method, as opposed to the classical evolutionary synthesis method in LH95.
Isochrone synthesis was introduced by Charlot \& Bruzual (1991) as a technique
to overcome the discrete appearance of the model predictions at late
evolutionary stages when the resolution in mass of the evolutionary models
becomes inadequate. Instead of binning stars in mass and assigning them to
a specific track, continuous isochrones are calculated by interpolating between
the tracks in the HRD on a variable mass grid using a subroutine originally
provided by G.~Meynet. This scheme generally produces smooth output products, 
the only exception being quantities related to supernovae (see below).

The {\em time resolution} of the model series is 0.1~Myr. Spectral energy 
distributions, however, are given only at time steps during which significant
changes occur.

The model predictions are discussed in the following sections. A summary of
all the figures is in Table~1. Each figure has five panels (a-e) for
each of the five metallicities. An exception are 
Figures~13 through 36 which show
UV line spectra. These spectra are for \Zs\ only, as the library stars are
currently available only for this metallicity. The models will be extended
as soon as non-solar library stars become available (Robert 1999).
The data files which were used for the
generation of the plots can be viewed and downloaded (see instructions on
the website). The files with spectra (Figures~1 
through 36) contain data at more time steps than plotted. 
For clarity reasons we included only a subset of all time steps in the plots. 
Only the figures for solar metallicity models are reproduced in the hardcopy 
version of the journal (panels `b' of Figures~1 through 13  and 37 through 120,
and Figures 13 through 36). The full data set for all metallicities is included
in the electronic version of the journal and can be accessed on our website
at {\tt http://www.stsci.edu/science/starburst99/}.

\section{Spectral energy distributions}

Spectral energy distributions for all five metallicities, the three
IMF parameterizations, and the two star-formation laws were calculated. The
wavelength coverage and spectral resolution are identical to those of
the model atmosphere set Lejeune et al. (1997). The minimum and maximum
wavelengths are 91~\AA\ and 160~$\mu$m, respectively. The spectral resolution
is wavelength dependent. It is typically about 10 to 20~\AA\ in the UV to
optical range. The energy distributions were calculated at intervals of
1, 10, and 100~Myr between 1 and 20~Myr, 20 and 100~Myr, and 
100 and 1000~Myr, respectively.

The standard IMF case ($\alpha=2.35$; \Mup~=~100~\Ms) is in 
Figure~1. Panels a-e show the energy distributions for 
the five considered metallicities. Not all time steps are plotted in the figures
but they are included in the ascii table file which can be downloaded. The
wavelength range has been restricted to 100~--~10,000~\AA\ in the figures,
but the full range is covered in the tables.

The most dramatic changes occur in the far-UV below 912~\AA, and even more 
below 228~\AA\ in the He$^{++}$ continuum. O stars dominate during the first 
few Myr, followed by a brief period at $\sim$4~Myr when WR stars contribute 
with their strong far-UV flux. Eventually the burst population fades away. 
Metallicity enters both via the continuum and line opacity (compare the
2600~\AA\ region at low and high metallicity) 
and via stellar evolution. The latter
effect can most readily be seen in the strength of the He$^{++}$ continuum.
It is strongest at high metallicity 
when WR stars preferentially form, and it is 
weakest at low metallicity (see also the discussion in \S~8).

The continuous star formation case for the same IMF is in 
Figure~2a-e. Most of the prior discussion holds for this
figure as well. Once the stellar types responsible for the photon contribution
to a wavelength interval have reached equilibrium between birth and death, 
the particular wavelength region becomes time independent.

We move on to different IMFs. The instantaneous and continuous cases, with
$\alpha=3.3$, \Mup~=~100~\Ms\ are in 
Figures~3a-e and 4a-e,
respectively. The truncated IMF,
$\alpha=2.35$, \Mup~=~30~\Ms\ is covered in 
Figures~5a-e and 6a-e. The
spectral energy distributions produce a softer radiation field as compared
with the standard IMF, in particular if stars with \Mup~$>$~30~\Ms\ are
suppressed.

The radiative properties shown later in the paper were obtained 
from these energy distributions. For instance, colors were derived by
convolving the spectra by the filter profiles. Users who wish to compute
colors in different filter systems can do so with the spectra in
Figures~1 through 6. Another application
would be to use the spectral energy distributions as input to a photoionization
code. In this case, Figures~1 through 6
should {\em not} be used because the nebular continuum was added to the
stellar flux. We computed an additional set of stellar spectra with exactly
the same parameters as before, but without including the nebular continuum.
This set may also be useful to compute magnitudes and colors for comparison
with objects which have no nebular continuum contribution. An example could
be a background subtracted star cluster, where the background subtraction
removed the nebular emission as well. There may also be cases where the
nebular emission of a galaxy or H~II region is more extended than the ionizing
cluster, and part of the nebular flux is outside the instrument aperture. 
Finally, the H~II region may not be ionization-bounded. Such cases may be
easier to model using energy distributions without nebular continuum, and
applying an empirical correction, e.g., from the observed \Ha\ flux in the
aperture. The models without nebular continuum
are in Figure~7a-e ($\alpha = 2.35$, \Mup~=~100~\Ms,
instantaneous star formation),
Figure~8a-e ($\alpha = 2.35$, \Mup~=~100~\Ms,
continuous star formation), 
Figure~9a-e ($\alpha = 3.3$, \Mup~=~100~\Ms,
instantaneous star formation),
Figure~10a-e ($\alpha = 3.3$, \Mup~=~100~\Ms,
continuous star formation),
Figure~11a-e ($\alpha = 2.35$, \Mup~=~30~\Ms,
instantaneous star formation), and 
Figure~12a-e ($\alpha = 2.35$, \Mup~=~30~\Ms,
continuous star formation). 
The spectra in Figures~7 through 12 and the curves in
Figures~45, 46, 75, and 76
are the only models in this paper which do not
include the nebular continuum.

\section{Ultraviolet line profiles}

Two spectral regions with useful diagnostic lines were computed at higher 
resolution. The wavelength range from 1200~\AA\ to 1850~\AA\ has the
strong resonance lines of \civ, \siiv, and \nv. At shorter wavelengths,
\ovi\ and Ly$\beta$ are of interest. The spectra described in this section
were computed with a normalized library of stellar UV spectra and absolute
fluxes derived from the energy distributions shown in the previous section.
We did not use the energy distributions directly since they are not featureless.
Rather we derived a featureless continuum by fitting a spline through 
line-free sections of the model atmospheres.

\subsection{The 1200~\AA\ to 1850~\AA\ region}

The 1200~\AA\ to 1850~\AA\ region is readily accessible to many satellites, like
IUE or HST. We synthesized spectra from an IUE high-dispersion library of
O- and WR-spectra and a low-dispersion for B stars. 
The spectral library and the method are discussed in Robert, Leitherer,
\& Heckman (1993) and Leitherer et al. (1995). 
The resolution is 0.75~\AA, adequate for typical spectral data
of extragalactic objects. This resolution is only reached during phases
dominated by O- and WR stars, as is the case for young ($t < 7$~Myr) bursts
and for the continuous case at any time. Otherwise the effective spectral
resolution approaches the resolution of IUE low-dispersion spectra (6~\AA). 
The upgrade of the B-star library from low- to high-dispersion has been
completed and will be implemented in the future (de Mello, Leitherer, \& 
Heckman, in preparation; Robert, Leitherer, \& Heckman, in preparation).

Only solar metallicity models are shown since the library stars have solar or
somewhat sub-solar metallicity.  A spectral library of massive stars in the 
Magellanic Clouds is being developed and will be available in the future
(Robert 1999; Robert, Leitherer, \& Heckman 1999, in preparation).

Figure~13 shows rectified spectra between 1 and 20~Myr
for an instantaneous burst. A standard IMF was used. The evolution of the
strongest {\em stellar-wind} features of \civ, \siiv, \nv, \niv, and \heii\ 
can be readily seen. These features disappear or become {\em photospheric}
lines after about 7~Myr when the transition from an O-star to a B-star
dominated population occurs. A purely photospheric line is, e.g.,
S~V $\lambda$1502.
The library stars have strong {\em interstellar} lines as well. One of the
strongest lines is \cii.

The continuous case for the same IMF is shown in 
Figure~14. The same basic effects are there as well
but the time-dependence is much weaker, and a quasi-equilibrium state is
reached after about 5~Myr. Figures~15 and 
16 show spectra with the same parameters as in 
Figures~13 and 14, but in
absolute luminosity units. 

As we did for the spectral energy distributions in the previous section, 
we re-computed the UV spectra for the two other IMFs, a steeper IMF with
$\alpha = 3.3$ and a truncated IMF with \Mup~=~30~\Ms. These spectra are
in Figures~17 through
24. The general trend is a weakening of all
stellar-wind lines, most notably \civ\ and \siiv, as these lines are due
to massive O stars. These lines are very sensitive IMF tracers. Their detection
immediately indicates an O-star population and therefore star-formation over
the past $\sim$10~Myr.

\subsection{The 1015~\AA\ to 1060~\AA\ region}

The wavelength region shortward of Ly$\alpha$ in local starburst galaxies
is accessible to, e.g., the FUSE mission (Sahnow et al. 1996)
and can be observed in star-forming
galaxies at high redshift from the ground. One of the strongest stellar-wind
lines is \ovi\ (Walborn \& Bohlin 1996). Synthetic spectra for the region
between 1015~\AA\ and 1060~\AA\ were presented by Gonz\'alez Delgado et al.
(1997) and were regenerated for the Starburst99 package. They are based on an 
empirical Copernicus library of O and B stars and have a resolution of 0.2~\AA.
A few library stars were observed with HUT at a spectral resolution
of about 3~\AA.
Only solar metallicity models are considered, as we have no library stars with
non-solar metallicity. 
Available time steps are 1 to 20~Myr at $\Delta t = 1$~Myr and
20 to 100~Myr at $\Delta t = 10$~Myr. Not every time step is plotted in the
figures.

We discuss the most notable features using the standard IMF, instantaneous
burst, rectified model (Figure~25). The strongest lines are
\ovi, Ly$\beta$, \ciia, and the Lyman and Werner bands of H$_2$, like the
one at 1050~\AA. In the first few Myr of the starburst, \ovi\ has a strong
P~Cygni profile, indicating winds from massive O stars. \ovi\ becomes weaker
with age and gradually blends with Ly$\beta$ at 1026~\AA. Ly$\beta$ is mostly
stellar, with some interstellar contribution. The line increases in strength
with increasing B-star fraction. \ciia\ follows Ly$\beta$ in its temporal
behavior. It is a strong B-star line with some interstellar contribution and
can be used to estimate the starburst age (Gonz\'alez Delgado et al. 1997). 
The continuous case with the same IMF parameters behaves accordingly
(Figure~26). The counterparts of 
Figures~25 and 26 in luminosity
units are in Figures~27 and 28.

The remaining figures for \ovi\ are organized the same way as the figures
for the 1200~\AA\ to 1850~\AA\ region. They are intended to highlight IMF
variations (Figures~29 through 36).
An IMF biased against massive O stars suppresses \ovi\ and strengthens \ciia.
Even at the earliest ages the absorption trough around 1025~\AA\ is mostly due
to Ly$\beta$, and not \ovi.

\section{Massive-star inventory}

In this section we give number predictions for massive stellar types that 
can be easily ``counted'' even in distant galaxies: O stars, WR stars, and
supernovae (SNe). 

O-star numbers are in Figures~37a-e and 38a-e 
for the instantaneous and continuous case, respectively. Panels a, b, c,
d, and e are for metallicities 0.040, 0.020, 0.008, 0.004, and 0.001,
respectively. All remaining figures in this paper have the same structure.
The O-star numbers include stars with spectral types O3 to O9.5 of 
all luminosity classes. The adopted spectral-type versus effective temperature
(\Teff) and luminosity ($L$) relation is from Schmidt-Kaler (1982). The
chosen star-formation histories produce up to about $10^3$ to $10^4$ O stars
at any time, depending on the age of the population. A flatter
IMF increases the O-star number.

WR stars are the evolved descendants of massive O stars. We define them as stars
with $\log T_{\rm{eff}} > 4.4$ and surface hydrogen abundance less than 0.4
by mass. The mass limits for the formation of WR stars are taken from 
Maeder \& Meynet (1994). WR stars have very extended atmospheres producing 
strong emission
lines which can be detected in distant starburst populations (Conti 1991).
The quantity of interest for comparison with models is the WR/O ratio, which
is plotted in Figures~39a-e and 40a-e for the
instantaneous and continuous case, respectively. This ratio is very metallicity
dependent: metal-rich starbursts are predicted to have more WR stars at any
time and to have longer phases when WR stars are present. This behavior
is generally in agreement with observations (Meynet 1995). 
The panels with
sub-solar 
metallicity have no graph for the IMF with \Mup~=~30~\Ms. This results
from the absence of WR stars at low metallicity. The detailed
shape of the graphs in Figures~39a-e and 40a-e
is very model dependent and future revisions of the evolutionary models may
change the WR/O ratio. Dedicated evolutionary synthesis models with 
particular attention
to the WR phase were published by Meynet (1995) and Schaerer \& Vacca (1998).
The binary channel to form WR stars in stellar populations was investigated 
by Cervi\~no \& Mas-Hesse (1997), Vanbeveren et al. (1997), Schaerer
\& Vacca (1998), and Dionne (1999).

WR stars come in two main subclasses: nitrogen-rich WN stars and carbon-rich
WC stars. We adopt the classification schemes of Conti, Leep, \& Perry (1983)
for WN stars and of Smith \& Hummer (1988) and Smith \& Maeder (1991) for WC 
stars. The WC/WN ratio measures the relative lifetimes spent in the two
phases (Figures~41a-e and 42a-e).
However, the calculations are affected by
  uncertainties in the interpolation of the relatively sparse tracks,
  as discussed in Schaerer \& Vacca (1998). In this respect, the present models
  are identical to those of Schaerer \& Vacca, who predict fewer WC stars than
  Meynet (1995) despite the use of the same tracks.
Most of the 
comments on the WR/O ratio apply to these figures as well. The WC/WN ratio
varies very little with the IMF exponent: the $\alpha=2.35$ and 3.3 cases
are almost indiscernible in the figures. This indicates that both stellar
types have progenitors of very similar mass. As a result, applying different
weights to different mass intervals does not affect the ratio. The same
argument applies to several other quantities discussed further below, like,
e.g., the CO index (Figure~95).

All massive stars with initial masses above 8~\Ms\ are assumed to explode as
SNe. This leaves open the question if a critical mass exists above
which stars directly form a black hole rather than producing a core-collapse
SN (Maeder 1992). If so, the predictions of 
Figures~43a-e and 44a-e would hardly be changed 
because the SN rate is strongly weighted towards the lowest
progenitor masses.

The user should be aware of a computational issue which can be seen in 
Figures~43a-e and 44a-e. The graphs show discontinuities
at the factor-of-2 level where there should be none (e.g., around 10~Myr in
Figure~43). A simple argument suggests that the SN rate
should be a smooth and almost constant function of time. The rate scales with
the product of the IMF and the mass-age relation. For a Salpeter IMF and
a mass-age relation $t(m) \propto m ^{-\gamma}$ with
$\gamma \approx 1.2$ (Schaerer et al. 1993b) the SN rate of an instantaneous 
burst becomes essentially time-independent (Shull \& Saken 1995). In other 
words, there are more and more SN events for lower
masses since there are more progenitors available, but
at the same time it takes longer for a star to turn into a SN. Obviously
the discontinuities in Figure~43 are not physical. They result
from a peculiarity of the isochrone synthesis interpolation technique. The
code attempts to optimize the mass interval size for the interpolation. This
works extremely well, except when only stars in an infinitesimal mass interval
are relevant, as is the case for the SN rate. The same applies to
all quantities which are directly dependent on the SN rate, like the
energy release from SNe. We decided not to artificially smooth the curves
since this would be a subjective process and we did not want to treat 
SN-related quantities different from the other predictions. We remind the
user to apply common sense when interpreting the figures. The correct 
supernova rate
between 5 and 35~Myr for an instantaneous burst in Figure~43 is 
about $10^{-3}$~yr$^{-1}$ (LH95).

\section{Luminosities}

We give the total and monochromatic luminosities at selected wavelengths. We
have chosen the $V$, $B$, and $K$  passbands, as well as a UV wavelength
at 1500~\AA\ since these are the most interesting cases for comparison
with observations. Luminosities for other bands can be obtained  from the 
spectral energy distributions in Figures~1 through 6.

The bolometric luminosity ($M_{\rm{Bol}}$) is defined such that the Sun has
$M_{\rm{Bol}} = 4.75$. The results in 
Figures~45a-e and 46a-e were obtained by integrating
the spectral energy distributions without the nebular continuum. The total
luminosity of the combined stellar and nebular continuum is smaller than the
pure stellar continuum since only a fraction of the absorbed stellar ionizing
flux is re-emitted in the nebular continuum. The curves
are smooth without strong discontinuities and reflect the behavior of
the most massive stars which provide most of the radiative energy output.

The absolute magnitudes $M_{\rm{V}}$, $M_{\rm{B}}$, and $M_{\rm{K}}$ are in 
Figures~47a-e (instantaneous) through 52a-e (continuous).
The absolute magnitude
$M_{\rm{V}}$ was calculated from the bolometric luminosity and the 
bolometric correction, which is 0.00 for a solar metallicity star with 
\Teff~=~7000~K and $\log g = 1.0$. The Sun has a bolometric correction of
$-0.19$ in this system. The absolute magnitudes
$M_{\rm{B}}$ and $M_{\rm{K}}$ follow from $M_{\rm{V}}$
and the $(B-V)$ and $(V-K)$ colors. All three luminosities have conspicuous
variations around 10~Myr when red supergiants (RSG) appear. The effect is
strongest in the $K$ band, which is closest to the energy peak of RSGs. The
RSG feature is very metallicity-dependent. The RSG contribution in 
Figure~51a-e is strongest at $Z=0.040$ and weakest at $Z=0.001$.
This apparent metallicity dependence results from the failure of the 
evolutionary models to predict correct RSG properties at $Z \leq 0.008$. This
effect was pointed out before by Mayya (1997) and Origlia et al. (1998).
Low-metallicity RSG models have too high surface temperatures and too short
RSG lifetimes. Uncertainties in the mixing processes at low 
metallicity are a possible explanation (Langer \& Maeder 1995). 
Currently there are no self-consistent stellar evolution models
available which correctly predict the variations of blue-to-red supergiants
with metallicity. Therefore our (and most other) synthesis models are incorrect
during phases when RSGs are important. An empirically adjusted set of synthesis
models was prepared by Origlia et al. (1998) but {\em these adjustments were not
made for the model set in this paper}.

Luminosities at 1500~\AA\ 
(Figures~53a-e and 54a-e) were computed by
averaging over the wavelength interval 1490~\AA~--~1510~\AA. This wavelength
becomes observable from the ground at redshifts larger than $\sim$2, and
the luminosity at 1500~\AA\ is a useful indicator of the star-formation rate,
or, in connection with the \Ha\ luminosity of dust obscuration (Pettini
et al. 1997). The 1500~\AA\ luminosity is a very robust prediction, with few 
uncertainties since the continuum at this wavelength comes from well-understood 
late-O/early-B stars.

\section{Colors}

Optical and near-infrared (IR) colors were calculated by convolving the
spectral energy distributions with the filter profiles. The filters are
in the Johnson (1966) system and are the same as in LH95. The zero point
is defined by a star with $Z=0.020$, \Teff~=~9400~K, and $\log g = 3.95$,
which has zero colors in all passbands. Colors in other photometric systems
can be obtained by convolving the spectral energy distributions with the
desired filter profiles.

These colors are available:
$(U-B)$ (Figures~55a-e and 56a-e),
$(B-V)$ (Figures~57a-e and 58a-e),
$(V-R)$ (Figures~59a-e and 60a-e),
$(V-I)$ (Figures~61a-e and 62a-e),
$(V-J)$ (Figures~63a-e and 64a-e),
$(V-H)$ (Figures~65a-e and 66a-e),
$(V-K)$ (Figures~67a-e and 68a-e), and
$(V-L)$ (Figures~69a-e and 70a-e).
As a reminder, the {\em continuous} nebular emission is included in the 
colors, but not the line emission. The $R$ filter is the most likely passband
to suffer from nebular line contamination since \Ha\ is included. A first check
of the expected degree of contamination can be made by comparing the \Ha\
equivalent widths of Figures~83a-e and 84a-e with
the width of the $R$ filter (about 2000~\AA). Line strengths of other
strong lines in H~II regions can be estimated from the photoionization
models of Stasi\'nska \& Leitherer (1996).

The RSG issue raised before applies to some of the color plots for low
metallicities as well. The strong metallicity dependence of the RSG feature
around $10^7$~yr is related to the effects discussed in \S~6.
 
Rather than colors, we give continuum slopes in the UV. We define the slope
$\beta$ as the spectral index of the spectral energy distribution:
$F_\lambda \propto \lambda^\beta$. Two indices are shown, one for the
average slope between 1300~\AA\ and 1800~\AA\ 
(Figures~71a-e and 72a-e), 
 and one for the 2200~\AA\ to 2800~\AA\ region
(Figures~73a-e and 74a-e). The slopes were
simply derived by fitting a first-order polynomial to the spectra
through the wavelength
intervals 1280~\AA~--~1320~\AA\ and 1780~\AA~--1820~\AA\ for $\beta_{1550}$,
and through 
2180~\AA~--~2220~\AA\ and 2780~\AA~--~2820~\AA\ for $\beta_{2500}$. This
should serve as an approximate guide for the variation of the slope with
time but becomes increasingly meaningless if the actual spectrum deviates
from a power law. Figures~1 through 6
suggest that spectral energy distributions with ages less than
$\sim$200~Myr are indeed well approximated by a power law in the UV but that
this assumption is no longer correct at older ages. The UV slopes are quite
independent of evolution and IMF effects for the first 30~Myr. This property
makes them very useful for deriving UV extinctions in galaxy spectra
(Calzetti, Kinney, \& Storchi-Bergmann 1994). Therefore the spectral slope of 
the restframe ultraviolet spectra of star-forming galaxies at high redshift
can be used to estimate the effects of dust obscuration in the early 
universe (Calzetti \& Heckman 1999, and references therein).

\section{Far-ultraviolet properties}

The predictions in this section rely on our capabilities to model the stellar
far-UV continuum below 912~\AA\ since this spectral region is generally not 
accessible to direct observations. A comparison between different models 
{\em for hot stars} in
this spectral region has been made by Schaerer \& de~Koter (1997). The
{\em internal} consistency, as judged from differences between the models,
is within 0.1~dex between 912~\AA\ and 504~\AA, which is relevant for the
flux ionizing neutral hydrogen (\nhi). Uncertainties become larger towards
shorter wavelengths, in particular below 228~\AA, where the emergent flux
becomes strongly dependent on stellar-wind properties. This of course leaves
open the question of the {\em external, absolute} uncertainties. Model
atmospheres generally do a good job above 228~\AA\ when combined with 
photoionization models and compared to H~II region spectra 
(Garc\'{\i}a-Vargas 1996) although there are some properties which are sensitive
  to the flux distribution in the neutral He continuum 
(Stasi\'nska \& Schaerer 1997). 
This makes errors in the integrated photon fluxes by more than 0.3~dex unlikely.
The region below 228~\AA\ has not yet been tested in such detail and the
uncertainties are potentially large.

Far-UV models for stars colder than about 30,000~K (types B and later) are
lagging behind. Cassinelli et al. (1995) found discrepancies by two orders
of magnitude between the observed extreme-UV flux of the B2 star
$\epsilon$~CMa and model predictions. Even slight errors in the adopted
wind parameters can produce huge far-UV flux variations at these relatively
low temperatures (Schaerer \& de~Koter 1997). 
 
We define the Lyman break as the ratio of the average flux in the wavelength 
interval 1080~\AA~--~1120~\AA\ over that between 870~\AA\ and 900~\AA. These
wavelength intervals are relatively free of line-blanketing so that we
measure mostly temperature rather than line opacity. The Lyman break is about 
a factor of 2 to 3 in O-star dominated phases and drops thereafter
(Figures~75a-e and 76a-e). The warnings about 
model uncertainties for the Lyman continuum of cooler stars apply to the
instantaneous case. Figure~75a-e should not be overinterpreted
after about 50~Myr since B stars like $\epsilon$~CMa (see previous
paragraph) could dominate. The Lyman break for a continuous population is 
always dominated by hot stars so that Figure~76a-e can be trusted
over the entire range plotted. 

The number of photons capable of ionizing neutral hydrogen (\nhi), neutral
helium (\nhei), and ionized helium (\nheii) were calculated by integration
of the spectra below 912~\AA, 504~\AA, and 228~\AA, respectively. They are
shown in 
Figures~77a-e and 78a-e (\nhi),
Figures~79a-e and 80a-e (\nhei), and
Figures~81a-e and 82a-e (\nheii).
As stated before, the numbers become increasingly uncertain with shorter
wavelength. \nheii\ is almost entirely produced by WR stars, and uncertainties
in the wind properties enter.  

We also give equivalent widths of several popular hydrogen recombination
lines. The continuum is taken from the model atmospheres and does not
include underlying stellar absorption.
The nebular continuum is of course taken
into account. The individual plots are for 
\Ha\ (Figures~83a-e and 84a-e),
\Hb\ (Figures~85a-e and 86a-e),
\Pab\ (Figures~87a-e and 88a-e), and
\Brg\ (Figures~89a-e and 90a-e). 
The transformation relations from \nhi\ to the line luminosities are in LH95.

\section{Other diagnostic lines}

In this section we present a few more diagnostics that can be useful to
isolate a stellar population in a galaxy spectrum. They are all related to
stars off the main-sequence: WR stars, RSGs and SNe.

In Figures~81 and 82 we predicted \nheii, which
can immediately be converted into the emission-line flux of {\em nebular}
He~II $\lambda$4686. Generally, a hot-star population capable of producing
nebular He~II will also show broad, {\em stellar} He~II $\lambda$4686 
(Schaerer \& Vacca 1998). Our model predictions for the stellar line are
in Figures~91a-e and 92a-e. The feature is commonly
referred to as the ``WR bump''. It includes only He~II and none of the other
nearby spectral features like C~III, N~III, [Ar~IV], and [Fe~III]. Note that
our code predicts other WR lines from the list of Schaerer \& Vacca but they
are not given here.

[Fe~II] $\lambda$1.26 is useful to count supernovae in starbursts
(Figures~93a-e and 94a-e). The supernova shock wave
destroys interstellar dust grains, thereby releasing iron atoms and ions
which had condensed on the dust grains. We adopted the scaling relation of 
Calzetti (1997) to convert the supernova rates into [Fe~II] line luminosities.

The remaining figures in this section are related to RSG properties. We 
repeat again that the predicted properties of 
post-main-sequence stars are not reliable and that all phases dependent on
RSG properties at sub-solar metallicity  are suspect. We begin with the 
spectroscopic 
CO index at 2.2~$\mu$m (Figures~95a-e and 96a-e).
We follow the definition of Doyon, Joseph, \& Wright (1994) who expressed
the CO strength as a function of temperature for dwarfs, giants, and
supergiants. The index is set to 0 for \Teff~$>$~6000~K. We also computed
the strength of the calcium triplet at $\lambda\lambda$8498~\AA, 8542~\AA, 
8662~\AA\ (Figures~97a-e and 98a-e). The equivalent
width of the sum of all three lines is related to gravity ($\log g$) and 
metallicity by
the relation $W(Ca T) = 10.21 - 0.95 \log g + 2.18 \log Z/Z_\odot$
(D\'{\i}az, Terlevich, \& Terlevich 1989; Garc\'{\i}a-Vargas, Moll\'a, \&
Bressan 1998). The predicted values neither take into account nebular 
emission nor stellar absorption of higher Paschen lines which fall in this 
wavelength region. Observations
must be corrected for these lines, if present, before a comparison is made.

Evolutionary synthesis models for the first and second overtones of
CO at 2.29~$\mu$m and 1.62~$\mu$m and for Si~I at 1.59~$\mu$m were presented
by Origlia et al. (1998). The models in
Figures~99a-e and 100a-e
(CO $\lambda$1.62~$\mu$m),
Figures~101a-e and 102a-e
(CO $\lambda$2.2~$\mu$m), and in
Figures~103a-e and 104a-e
(Si~I $\lambda$1.59~$\mu$m) are expanded versions of the {\em unmodified}
case discussed by Origlia et al. In that paper, the effect of modifying the
evolutionary tracks was studied and several sets of IR-line models were
published. To be consistent with the other quantities shown here, we opted
for including only the standard models. The models for
Si~I $\lambda$1.59~$\mu$m were not covered by Origlia et al. and are shown
here for the first time. The theoretical library of Origlia et al. (1993)
which was used for Si~I is less reliable than that for CO so that care is
required when using the Si~I models.

\section{Mass and energy return}

The previous sections cover stellar numbers and radiative properties. Here we
turn to non-radiative properties of the stellar population. The input physics
is discussed in greater detail in Leitherer et al. (1992). The figures in this
section show the mass and energy input by stars and SNe. Only 
core-collapse SNe are considered. They are assumed to release 
$10^{51}$~erg per event in the form of kinetic energy, independent of
metallicity. We do not address the efficiency of thermalization, which
would require hydrodynamical modeling. The simulations of Thornton et al.
(1998) suggest that about 10\% of the available kinetic energy can actually
be used to pressurize the interstellar gas. The remaining 90\% are radiated
away.

The rate of mass return of stellar-wind and SN material is plotted in 
Figures~105a-e and 106a-e. The adopted mass-loss rates
are {\em not} those of the evolutionary models but those favored by Leitherer
et al. (1992) and LH95. We prefer this approach over simply using the 
evolutionary mass-loss rates. In our opinion, the evolutionary mass-loss rates
are too high (see Section~2) and should be considered only as an adjustable 
parameter in evolution models but are not directly related to the observed
mass-loss rate. Slight differences between Figures~105 and 
106 and the corresponding figures in LH95 are not due to a
different mass-loss parameterization but because of different stellar
parameters (\Teff, $L$) in the revised tracks. The individual contributions
from stellar winds and SNe are broken down in 
Figures~107a-e and 108a-e for the standard
IMF case. Stellar winds are generally more important for young bursts whereas
SNe take over at later times. 

The total mass return from winds and SNe is in
Figures~109a-e and 110a-e.
The quantity plotted is $\int \dot{M} dt$, i.e. the integral of the curves
in Figures~107 and 108 over time. This 
quantity is useful to evaluate the exhaustion of the gas supply in a galaxy
or the degree of chemical pollution by wind and supernova material.

Figures~111a-e and 112a-e give the mechanical
luminosity \Lm\ released by winds and supernovae. The curves are similar to
those for the mass return in Figures~105 and 106.
The relative contributions to \Lm\ from stellar winds and SNe are in 
Figures~113a-e and 114a-e. A further
break-down into individual stellar-wind components is given in Leitherer
et al. (1992). Generally, most of the wind power comes from WR stars, with
some contribution from O stars. All other stellar phases are negligible since
wind velocities of cool stars are lower by two orders of magnitude. The
energy return ($\int L_{\rm{mech}} dt$) from winds and SNe is in
Figures~115a-e and 116a-e.

It is instructive to perform a differential comparison between the radiative
and the non-radiative energy output of young stellar populations. This is
done in 
Figures~117a-e and 118a-e for the ratio
of the ionizing ($<$912~\AA) over the bolometric luminosity, and in
Figures~119a-e and 120a-e for the ratio
of the mechanical (\Lm) over the bolometric luminosity. The non-radiative energy
input into the interstellar medium becomes significant in comparison with
ionizing radiation once the strong WR winds are turned on. For a single
population, non-thermal and ionizing energy input become equally important
around $\sim$10~Myr.

\section{Conclusions}

We have computed a large grid of predictions for observable properties of
galaxies with active star formation. The distribution of the models is purely
web-based. We believe the community will find this method more useful than
a hardcopy publication.
All the figures discussed in this paper are at 
{\tt http://www.stsci.edu/science/starburst99/}.
This webpage provides links to other spectrophotometric databases as well.

It is worthwhile to recall the most important short-comings and uncertainties
of the models:\\
{\em Chemical evolution} is not treated self-consistently. Each stellar
generation has the same chemical composition. This becomes a concern for models
which are evolved over times during which significant changes in the
metallicity of the ISM occur. In this case spectra will have a wavelength
dependent metallicity. Generally, light at shorter wavelengths is produced
by more massive and younger stars, which are chemically more evolved than
older stars.\\
{\em Binary evolution} has been neglected. Although about 50\% of all 
stars form in
binaries, it is not clear if a significant fraction of these binaries has
an evolutionary history that differs substantially from single-star evolution.
There is disagreement in the literature on this point. Under these circumstances
we took the approach of implementing the simpler of the two alternatives, as
nature itself often prefers to do.\\
{\em Mass loss and mixing} processes in stellar evolution are still poorly
understood. Stellar phases, like WR stars or RSGs, are particularly affected
by such uncertainties. The main culprit is the lack of a self-consistent
theory which makes the introduction of adjustable parameters necessary. These
parameters then are sometimes extrapolated into a regime for which they
were not calibrated, such as metallicity.\\ 
The situation is particularly disturbing for {\em red supergiants} whose
properties are badly reproduced at low metallicity. There is no easy 
work-around for the user of evolutionary models, except for an empirical 
adjustment of the tracks. We have discussed such an adjusted model set. Clearly,
a strong effort on the stellar evolution modeling side is called upon for
improvements.\\
Our models put most of the emphasis on early evolution phases. {\em Later 
phases}, like AGB stars or white dwarfs are covered only
crudely or not at all. While improvements to our code can be made, our prime
goal and expertise is related to massive, hot stars, and we decided to 
optimize this stellar species first.

Despite the warnings, the model set should turn out to be useful for
the interpretation of observations of star-forming galaxies. For maximum
benefit, the user is encouraged to compare our model predictions with those
of other groups, such as those mentioned earlier in the paper.

To provide maximum flexibility, we offer the user to run tailored models
at our website. Instructions on how to run the code and how to access the
results are given at the website. The Fortran code 
is distributed freely and can be retrieved from the website as well.

\acknowledgments

Harry Payne helped us trouble-shoot numerous bugs and pitfalls we encountered
during the construction of our webpage.
D. Foo Kune acknowledges financial support from the STScI Summer Student
Program. Salary support for D.~Devost and D.~Schaerer came from
the STScI Director's Discretionary Research Fund. We appreciate advice on
electronic publishing and data maintenance from Bob Hanisch.
Partial support for this work
was provided by NASA through grant number NAG5-6903, from the
Space Telescope Science Institute, which is operated by the Association
of Universities for Research in Astronomy, Inc., under NASA contract
NAS5-26555.

\clearpage

\normalsize

\end{document}